\documentclass[apj]{emulateapj}
\usepackage{hyperref}
\submitted{Accepted for publication in ApJ, May 18 2006}

\newcommand{\spi}{_{\mathrm{spine}}}
\newcommand{\rel}{_{\mathrm{rel}}}
\newcommand{\sh}{_{\mathrm{sheath}}}
\begin{document}

\title{New Chandra observations of the jet in 3C\,273.\\ I. Softer
  X-ray than radio spectra and the X-ray emission mechanism}

\author{Sebastian Jester\footnote{Presently Otto Hahn Fellow of the
  Max-Planck-Gesellschaft at the Department of Physics and Astronomy,
  University of Southampton, Southampton SO17 1BJ, UK}}
\affil{Particle Astrophysics Center, Mail Stop 127, Fermi National Accelerator Laboratory, PO Box
  500, Batavia, IL 60510}
\email{jester@phys.soton.ac.uk}

\author{ D.~E. Harris}
\affil{Smithsonian Astrophysical Observatory, 60 Garden St., Cambridge, MA
  02138}
\email{harris@head.cfa.harvard.edu}

\author{Herman L. Marshall} 
\affil{MIT Kavli Institute for Astrophysics and Space Research, 77
  Massachusetts Avenue, Cambridge, MA 02139}
\email{hermanm@space.mit.edu}

\author{Klaus Meisenheimer}
\affil{Max-Planck-Institut f\"ur Astronomie, K\"onigstuhl 17, 69117
  Heidelberg, Germany}
\email{meise@mpia.de}

\begin{abstract}
The jet in 3C\,273 is a high-power quasar jet with radio, optical and
X-ray emission whose size and brightness allow a detailed study of the
emission processes acting in it. We present deep Chandra observations
of this jet and analyse the spectral properties of the jet emission
from radio through X-rays.  We find that the X-ray spectra are
significantly softer than the radio spectra in all regions of the
bright part of the jet except for the first bright ``knot~A'', ruling
out a model in which the X-ray emission from the entire jet arises
from beamed inverse-Compton scattering of cosmic microwave background
photons in a single-zone jet flow.  Within two-zone jet models, we
find that a synchrotron origin for the jet's X-rays requires fewer
additional assumptions than an inverse-Compton model, especially if
velocity shear leads to efficient particle acceleration in jet flows.
\end{abstract}

 
\section{Introduction}
\label{s:intro}

Since the launch of the Chandra X-ray observatory, it has become
evident that X-ray emission is a common feature of jets in radio
galaxies and quasars (see, e.g., \citealt{WBH01} for low-power radio
galaxies, \citealt{SGMea04} and \citealt{MSLea05} for X-ray surveys of
powerful radio galaxies and quasars, the overview articles by
\citealt{HK02} and \citealt{KS05} for emission mechanisms, and the
XJET home page\footnote{\url{http://hea-www.harvard.edu/XJET/}} for an
up-to-date list of X-ray emission associated with extragalactic jets).
Typically, the X-ray emission from low-power jets (\citealp{FR74} [FR]
class I) fits on a single synchrotron spectrum with their radio and
optical emission and is thus satisfactorily explained as synchrotron
emission (with the interesting problem of having to accelerate X-ray
emitting synchrotron electrons \emph{in situ}).  However, the spectral
energy distribution (SED) of high-power (FR II) jets usually shows the
so-called ``bow-tie'' problem, i.e., their X-ray spectrum does not fit
on an extrapolation of the observed radio/optical synchrotron
spectrum.  A representative case in point is the first new X-ray jet
detection by Chandra, PKS~0637-752 \citep{Sch00,Cha00}, with an
observed cutoff to the synchrotron emission in the optical range, but
a fairly hard X-ray spectrum\footnote{We define spectral indices such
  that $S_\nu \propto \nu^\alpha$.} ($\alpha \approx -0.7$) and an
X-ray flux per frequency decade (i.e., $\nu S_\nu$) exceeding the
optical one.  As the predicted X-ray intensity from synchrotron
self-Compton (SSC) emission is usually orders of magnitude below the
observed one, it has been suggested that these jets are still highly
relativistic at kiloparsec scales, with bulk Lorentz factors in the
range 5--30 and beyond, and that the X-rays are inverse-Compton (IC)
scattered cosmic microwave background (CMB) photons \citep[the beamed
  IC-CMB process;][]{Tav00,Cel00}.  The high Lorentz factors are
necessary in order for the energy density of the CMB photon field to
be boosted sufficiently in the jet rest frame (by a factor $\Gamma^2$,
where $\Gamma$ is the \emph{bulk} Lorentz factor of the jet fluid) to
account for the observed X-ray:radio flux ratios.

The beamed IC-CMB process is generally invoked to account for the
X-ray emission from jets in which a single synchrotron component
cannot account for the emission in all wavebands.  If correct, the
inferred values of $\Gamma$ would constitute the first firm
measurement of the velocity of high-power jets \citep[for the
  velocities of low-power jets, see the modeling
  by][]{LB04,CL04,CLBea05} with profound implications for the analysis
of jet energetics and composition.  Since the increase of the CMB
energy density with redshift would cancel the surface brightness
dimming with redshift, beamed IC-CMB X-ray jets at high redshift could
be cosmic beacons that outshine their quasars \citep{Sch02}.
Moreover, there \emph{are} jets in FR~II radio galaxies and quasars
whose X-ray emission is either well explained as synchrotron emission
on an extrapolation of the radio-optical spectrum (e.g., in some of
the jets in the survey by \citealt{SGMea04}, and that in 3C\,403
[\citealp{KHWM05}]), or because the necessary Doppler factors cannot
be achieved from geometrical considerations, such as in Pictor~A
\citep{HC05} where there is X-ray emission from the counter-jet as
well as the approaching jet.  A detailed check of the beamed IC-CMB
model is therefore warranted.

There are two basic testable predictions of the beamed IC-CMB model
which arise from the fact that those electrons responsible for the
upscattering of CMB photons into the X-ray band have low Lorentz
factors ($10 \la \gamma \la 100$) and hence emit synchrotron radiation
at wavelengths of a few tens or hundreds of Megahertz, below the
typical Gigahertz-range observing frequencies allowing high-resolution
studies of radio jets.  Essentially, a similar part of the electron
energy distribution can be observed in two different wavebands
\citep[the electrons producing IC emission in the Chandra band emit
  synchrotron radiation in the range 40\,Hz--1\,MHz;][]{HK02}, leading
us to expect a close correspondence both of morphology and of
spectral index between a beamed IC-CMB jet's X-ray and radio emission.

We have chosen the jet in 3C\,273 for a detailed test of the beamed
IC-CMB model with the Chandra X-ray Observatory.  This jet has a
unique combination of properties that make it worthy of detailed
study: it has a large angular size (bright radio, optical and X-ray
emission being observed from 11\arcsec\ out to 22\arcsec\ from the
core), it is one of the closest high-power jets ($z=0.158$) so that
smaller physical scales\footnote{We use a flat cosmology with
  $\Omega_{\mathrm{m}}=0.3$ and $H_{0} =
  70\,\mathrm{km\,s^{-1}Mpc^{-1}}$, leading to a scale of $2.7
  \mathrm{kpc}$ per second of arc at 3C\,273's redshift of 0.158.} can
be resolved than in most similar jets, and high-resolution VLA and HST
data are available \citep[][and references therein]{JRMP05} that allow
the construction of detailed SEDs and the study of the relation
between the jet's X-rays and the radio-optical synchrotron emission.
3C\,273's jet is among those from which X-ray emission had been
detected already with the \emph{Einstein} satellite
\citep{Wil81,HS87}.  Based on a comparison of ROSAT HRI observations
with radio and optical data, \citet{Roe00} concluded that the X-ray
emission could only be due to the same synchrotron component as the
jet's radio-optical emission in the first bright feature of the jet
(``knot~A''), but not in the remainder of the jet, where there is a
cutoff to the synchrotron component in the near-infrared/optical
range; later, \citet{Jes02} and \citet{JRMP05} showed that at 0\farcs3
resolution, already the radio-optical SEDs of most of the jet,
including knot~A, require a two-component model to account for the
emission.  Hence, the X-ray emission there cannot be due to a single
radio--optical--X-ray synchrotron component, either \citep[but
  see][who suggests that an additional contribution of synchrotron
  emission from particles moving in fields that are tangled on small
  scales may explain the observed flattening, at least in parts of the
  jet]{Fle05}.  \citet{Jes02} suggested that the second optical
component may be the same component as the jet's X-ray emission, but
could not constrain the emission mechanism for the high-energy
component further.

Regarding the X-ray emission mechanism, \citet{Roe00} ruled out SSC as
well as thermal Bremsstrahlung.  The first Chandra observations of
this jet were presented by \citet{Sam01} and \citet{Mareta01}.
\citet{Mareta01} presented the first X-ray image of this jet with both
high resolution and high signal-to-noise ratio.  Regarding the SEDs,
they came to similar conclusions as \citet{Roe00}. \citet{Sam01} had
analyzed a smaller set of the early Chandra data and favored a beamed
IC-CMB model for the emission from all parts of the jet.
\citet{Mareta01} compared the optical and X-ray morphology at the
Chandra resolution of 0\farcs78 and found that the X-ray and optical
emission come from the same parts of the jet and show very similar
features.  However, there are offsets between the X-ray and optical
peaks in one or both of the first two bright knots. In the remainder
of the jet, the relative variations in X-ray brightness are smaller
than those in the radio and optical.  Such size and brightness
differences require some parameter fine-tuning in the IC-CMB model.

Thus, previous observations had left the nature of the X-ray emission
mechanism of 3C\,273 unclear.  Here, we present a spectral analysis of
our new Chandra observations, which we combine with archival data and
our VLA + HST dataset from \citet{JRMP05}.  Observations and data
reduction are described in \S\ref{s:obs}, our spectral analysis and
its results are presented in \S\ref{s:analysis}, and we discuss in
\S\ref{s:discussion} the implications of our results for the beamed
IC-CMB model and other emission mechanisms. We summarize our findings
in \S\ref{s:conc}.

\section{Observations and data reduction}
\label{s:obs}

We have obtained four observations of 3C\,273 and its jet in Chandra
observing cycle 5, totaling just under 160~ksec observing time.  The
observations were split into four exposures of equal length in order
to search for variability within the observing cycle.  The quasar and
jet were placed on the ACIS-S3 chip in each of these.  We also use
calibration observations of 3C\,273 from the Chandra archive: two
ACIS-S3 data sets from cycle~1, with 30~ksec exposure time each, and
seven ACIS+HETG calibration observations, totaling just under
200~ksec.  The gratings only transmit about 10\% of the soft X-ray
photons ($E < 2$\,keV), but about 75\% of the hard X-ray photons ($E >
6$\,keV).  Hence, we use the grating data to perform consistency
checks on the spectral shape of the jet at the high-energy end of the
Chandra bandpass.  We will not use them for morphological analysis.
This decision is based purely on the fact that their inclusion would
have required substantial additional effort without a commensurate
gain in the total number of detected photons; as the calibration data
were used by the calibration team to check the effective area, not the
pileup or contamination correction, they could be used without causing
any circular reasoning. Table~\ref{t:obslog} gives the observing log.
\begin{deluxetable}{lrrrr}
\tablecaption{Observing log for ACIS observations of 3C\,273.\label{t:obslog}}
\tablehead{
\colhead{Grating} & \colhead{ObsId} & \colhead{Exp. time} & \colhead{Obs. start time } & \colhead{Frame time} \\
& \colhead{} & \colhead{ksec} & \colhead{UT} & \colhead{sec}
}
\startdata
None & 1711 & 28.09 & 2000-06-14 05:13:19 & 1.1 \\
& 1712 & 27.80 & 2000-06-14 13:43:27 & 3.2 \\
& 4876 & 41.30 & 2003-11-24 23:08:40 & 0.4 \\
& 4877 & 38.44 & 2004-02-10 03:40:39 & 0.4 \\
& 4878 & 37.59 & 2004-04-26 20:55:16 & 0.4 \\
& 4879 & 39.23 & 2004-07-28 03:35:33 & 0.4 \\
\hline
HETG & 459 & 39.06 & 2000-01-10 06:46:11 & 2.5 \\
& 2463 & 27.13 & 2001-06-13 06:41:21 & 2.1 \\
& 3456 & 25.00 & 2002-06-05 10:03:12 & 1.9 \\
& 3457 & 25.38 & 2002-06-05 17:19:11 & 2.5 \\
& 3573 & 30.16 & 2002-06-06 00:43:51 & 2.5 \\
& 4430 & 27.60 & 2003-07-07 12:08:58 & 3.2 \\
& 5169 & 30.17 & 2004-06-30 12:39:18 & 2.5
\enddata
\tablecomments{All grating-free observations used the ACIS-S3
  chip. ObsIds 1711 and 1712 and all HETG observations are from
  calibration programs, ObsIds 4876-4879 were obtained under Chandra
  GO proposal 05700741.}
\end{deluxetable}

We used the Chandra analysis software CIAO v.~3.2 or later and
calibration database CALDB v.~3.0 or later to reprocess all data,
taking advantage of updated calibrations, in particular the ACIS
contamination correction \citep{MTGea04}, and to remove the pixel
randomization.  The spectral analysis (see following section) was
performed using spectral fits in the Sherpa software package
\citep{FDS01}.  In addition, we constructed flux-calibrated brightness
maps for the ACIS-S3 observations.

Registering Chandra data sets is challenging. The absolute pointing
calibration of Chandra of about 0\farcs4 \citep[Table 5.1]{POG} is not
sufficient for our analysis, since the jet width is only about
1\arcsec.  The quasar core is the only bright X-ray point source in
the vicinity of the jet, but it is so bright that it is severely piled
up \citep[Section 6.14]{POG} even with a frame time of 0.4\,s (chosen
in our new observations) or with the HETG in place.  However, the
readout streak in the grating-free observations contains so many
counts that it can be used to constrain the location of the quasar
core in the perpendicular direction with high precision and accuracy.
The unpiled wings of the PSF provide the necessary second constraint.
We further attempt to mitigate the effect of pileup by constructing an
``eV per second'' (evps) map, an energy-weighted count rate map.
Three of the authors (SJ, DEH, HLM) have performed independent fits of
the non-piled up part of the PSF and the readout streak, both on the
evps maps and on the brightness maps, to obtain the position of the
quasar core in each of the six ACIS-S3 sets.  From the RMS of the
individual determinations, we estimate that the alignment is better
than 0\farcs1 in each coordinate.  We average the individual
measurements to get a final quasar position in ``physical'' ACIS
coordinates. We then update the world coordinate system (WCS) of the
individual event files and maps to assign 3C273's right ascension and
declination as determined from radio observations to the quasar's
fitted physical coordinates.  For the HETG zeroth-order images, which
are used for consistency checks, we fit a Gaussian to the wings of the
quasar image to fix the astrometry relative to the quasar core.

\section{Analysis and results}
\label{s:analysis}

\subsection{Source extraction regions, variability search, and spectral fits}
\label{s:analysis.fits}

\subsubsection{Extraction regions}
\label{s:analysis.fits.regions}

\begin{figure}
\plotone{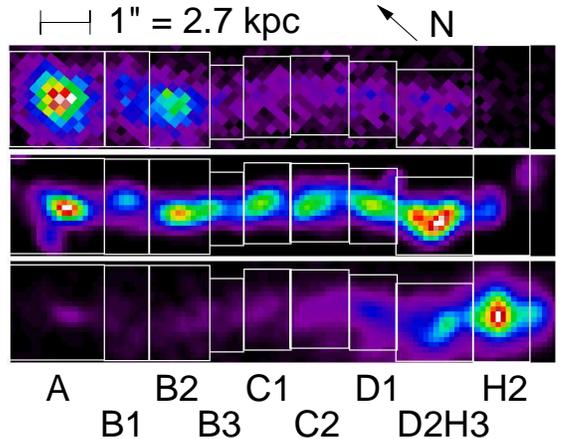}
\caption{\label{f:xraymap}The jet in 3C273 observed with Chandra
  (top), HST ($\lambda=620$\,nm, middle) and VLA ($\lambda=$3.6\,cm,
  below).  The long edges of each image are along a position angle of
  222\degr; the physical size of the region covered by each image is
  5.7\,kpc by 28.6\,kpc. The boxes indicate the regions adopted for the
  analysis in this paper. The VLA and HST map and the labeling scheme
  are taken from \protect\citet{JRMP05}, with a resolution of 0\farcs3
  (810\,pc) and a pixel size of 0\farcs1 (270\,pc). The Chandra map is
  from ObsId 4876, with counts binned in 0\farcs166 (450\,pc) pixels
  and a resolution of 0\farcs78 FWHM (2.1\,kpc).}
\end{figure}
We perform a separate spectral analysis in each of nine regions, which
we have defined considering both the radio/optical and X-ray
morphology.  Figure~\ref{f:xraymap} shows the regions on a Chandra,
HST and VLA image.  The nomenclature for the subdivisions of the
original regions ABCD \citep{LNHea84} is that established by
\citet[their Table 4]{RM91}; later, \citet{Baheta95} and thence
\citet{Mareta01} used different subdivisions.\footnote{The
  correspondence for the labels that are different between the former
  and the latter nomenclature is: A=A1; B1=A2; B2=B1; B3=B2; D1=C3;
  D2=D.}  There are no gaps between the extraction regions, and each
is about 1-2\arcsec\ in size.

While the jet shows similar morphology at all wavelengths, there are
differences in some places, which create some ambiguity as to where
the boundaries between different jet regions (or knots) should be
placed.  The differences in morphology at different wavelengths imply
spatial variations in the spectral shape.  Hence, some of our
measurement regions mix emission from jet features with different
spectral shapes.  Given that the Chandra resolution of 0\farcs78
(full-width at half-maximum intensity of a Gaussian fitted to the
readout streak) is comparable to the typical diameter of the jet
features, such mixing of spectral shapes cannot be avoided entirely.
However, the morphology differences are mostly minor, the sole
exception being region B1, in which the X-ray and optical emission
clearly arch to the north, while the radio emission has an arc to the
south \citep[compare][]{Mareta01}.  This difference may in fact
contain a crucial hint to the nature of the emission and particle
acceleration mechanism acting in the jet; however, the two arcs are
not fully resolved by Chandra, so that we include both in a single
region.  Still, the need to consider the emission from the arcs in B1
separately should be borne in mind when analyzing the spectral shape
there.

For the estimation of the background count rate, we define annuli of
similar radial extent as the corresponding jet region that are
centered on the quasar, as large-angle scattered photons from the
quasar core are the dominant contribution to the background.  The
background regions exclude the jet regions themselves as well as the
readout streak.

Considering the locations of those events classified as ``afterglow
events'' by the CIAO pipeline suggests that ObsIds 1711 and 1712
include some piled-up events at the brightness peak of region A. The
other ACIS-S3 observations used a 1/8th subarray, shortening the frame
time correspondingly, and therefore do not suffer from pileup.  The
HETG observations have lower count rates in the jet and pileup is much
less than 1\% there.

To determine the flux and X-ray spectral index, we extract spectra and
response files for each source region using the \texttt{psextract}
CIAO script.  

\subsubsection{Variability search}
\label{s:analysis.fits.variability}

We searched for variability in the count rates of each of the regions
over all six ACIS-S3 data sets and found no variations in excess of
the expected Poisson fluctuations for a constant count rate.  In
addition, we performed separate Sherpa power-law fits for each knot
and each epoch, as detailed below.  We used the $\chi^2$ test to
assess the significance of the fluctations.  The observed fluctuations
and their probabilites are given in Table~\ref{t:variab}.  In nearly
all regions, the $\chi^2$ test returned very high probabilities for
obtaining the observed fluctuations in the fit parameters across the
six epochs as random fluctuations of an underlying spectrum with both
constant flux and spectral index. Only the flux variability for
region~D2H3 is significant at the 96\% confidence level; however, at
3\%, it is also amongst the smallest observed fluctuations, and there
is no systematic trend.  Thus, there is no detection of significant
systematic variability in any of the regions.
\begin{deluxetable}{lrrrr}
\tablecaption{\label{t:variab}Results of variability search}
\tablehead{
\colhead{Region} &
\colhead{Flux RMS\tablenotemark{a}} &
\colhead{$p$\tablenotemark{b}} &
\colhead{Spectral RMS\tablenotemark{c}} &
\colhead{$p$\tablenotemark{b}}\\
&\colhead{\%} & & \colhead{\%} & }
\startdata
A &  6 &  0.92 &  4 &  0.90 \\ 
B1 &  9 &  0.99 &  8 &  0.88 \\ 
B2 &  7 &  1.00 &  2 &  0.08 \\ 
B3 &  5 &  0.08 & 11 &  0.49 \\ 
C1 &  6 &  0.35 &  9 &  0.58 \\ 
C2 &  4 &  0.10 &  7 &  0.40 \\ 
D1 &  9 &  0.75 &  9 &  0.51 \\ 
D2H3 &  3 &  0.04 &  6 &  0.39 \\ 
H2 & 14 &  0.12 & 18 &  0.16 \enddata
\tablecomments{Variability of jet features fitted with a power-law
  model as in \S\,\protect\ref{s:analysis.fits.model}}
\tablenotetext{a}{Root-mean-squared variability of power-law flux normalisation, relative to the mean}
\tablenotetext{b}{$\chi^2$ probability of obtaining the observed RMS variability
  at random, given the measurement errors}
\tablenotetext{c}{Root-mean-squared variability of power-law spectral index, relative to the mean}
\end{deluxetable}

In the remainder of this paper, we therefore concentrate on the
question whether the spectral shape of the jet emission is consistent
with the beamed inverse-Compton model.  The morphology will be
discussed in detail in a separate paper.

\subsubsection{Spectral fits}
\label{s:analysis.fits.model}

For the analysis in the remainder of the paper, we combine all six
epochs of ACIS-S data.  In each region, we perform a joint Sherpa
spectral fit for data and background, using simple power laws to
describe both the source and the background components.  As the quasar
itself shows both spectral and flux variability in the soft X-ray band
\citep{McHLNea99}, we allow an independent background model component
for every epoch.  We include a \texttt{JDPileup} model \citep{Dav01}
for the fit of region~A in ObsIDs 1711 and 1712; the use of this model
is appropriate since this region is unresolved by Chandra
\citep{Mareta01,MJHea05} and it is justified by the fit results which
indicate a pileup fraction of 3.6\% and 9.4\%, respectively. Trial
fits including a pileup model for the other 4 data sets and for all
other jet regions had results consistent with 0 piled up events, even
though scaling the piled up fraction by the frame time would lead us
to expect a pileup fraction of about 1\% (see frame times in
Table~\ref{t:obslog}).  As in \citet{Mareta01}, we use the hydrogen
column density of $n_{\mathrm{H}}=1.71\times 10^{20}$\,cm$^{-2}$ given
by \citet{ABMea93}.  We perform fits using two methods. First, we used
the likelihood method \citep[Cash statistic;][]{SherpaRefstats} for
unbinned events, using only the energy range
0.5~keV--8.0~keV. Secondly, we grouped the data to a minimum of 15
counts per bin with $\chi^2$ Primini statistics
\citep{SherpaRefstats}, in this case excluding all bins below 0.5~keV.
We do separate fits for the data from the HETG zeroth-order images and
from the grating-free observations.  The fit results from both data
sets (HETG zeroth-order images and grating-free S3 images) and both
fitting methods (unbinned/Cash and grouped/$\chi^2$) are consistent
with each other.  We have carried out further spot checks using XSPEC
power-law fits of the grouped data sets, in which background events
are subtracted before fitting (instead of being included in the fit as
a separate model component), and found excellent agreement with the
Sherpa results. Thus, the following analysis will make use of the fit
parameters obtained from the fits of unbinned data with Cash
statistics, and in the remainder of the paper, we only bin data for
display purposes.

\subsection{Fit results and SEDs from radio to X-rays}
\label{s:ana.res}

\begin{deluxetable}{lrrrrr}
\tablecaption{\label{t:fits}Fit results for joint X-ray spectral fits}
\tablehead{
\colhead{Region } &\colhead{Net
    counts\tablenotemark{a}} & \colhead{$\alpha$ } &
\colhead{$\sigma_\alpha$ } &\colhead{$S_\nu(1\,\mathrm{keV})$ } &
\colhead{$\sigma_S$} \\
& &  & & \colhead{nJy} & \colhead{nJy}}
\startdata
   A & 10849 & $-$0.83 & 0.02 & 46.54 & 0.54\\
  B1 &  2632 & $-$0.80 & 0.03 & 10.89 & 0.25\\
  B2 &  3992 & $-$0.97 & 0.03 & 19.98 & 0.33\\
  B3 &   768 & $-$1.13 & 0.07 & 3.41 & 0.14\\
  C1 &  1115 & $-$1.07 & 0.06 & 4.85 & 0.16\\
  C2 &  1467 & $-$0.96 & 0.05 & 6.25 & 0.18\\
  D1 &  1182 & $-$1.02 & 0.05 & 5.16 & 0.17\\
D2H3 &  1707 & $-$1.04 & 0.04 & 7.82 & 0.20\\
  H2 &   317 & $-$1.27 & 0.12 & 1.30 & 0.09
\enddata
\tablecomments{The fits were done on the six ACIS-S3 data sets without
  grating, using original events in the range 0.5~keV--8.0~keV with
  Cash statistics in Sherpa. Spectral indices are defined such that
  $S_\nu \propto \nu^\alpha$.}
\tablenotetext{a}{Background-subtracted number of counts summed over all 6
  data sets}
\end{deluxetable}
\begin{figure*}
\plotone{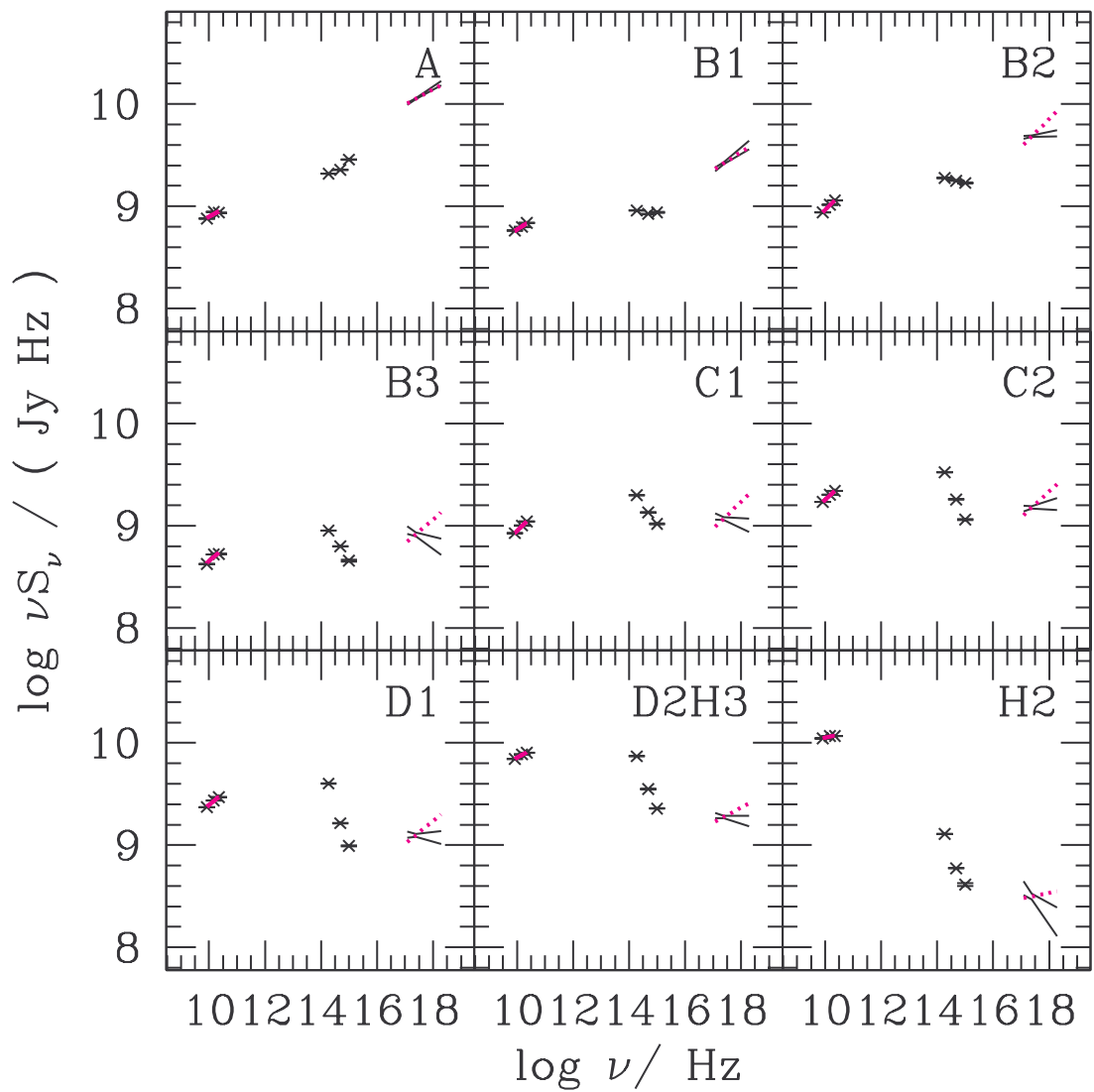}
\caption{\label{f:SEDs}Spectral energy distributions for the jet
  regions defined in Fig.~\protect\ref{f:xraymap}. The radio and
  optical points are obtained by integrating the flux from the maps at
  0\farcs3 resolution from \protect\citet{JRMP05} over the same
  extraction regions as the X-rays.  Error bars for radio and optical
  data show random errors; there may be additional calibration and
  flat-fielding errors of up to 3\% for the near-infrared and optical
  data points. The magenta solid line in the radio region is a least-squares
  power-law fit to the three VLA data points; the dotted line in the
  X-ray band has the same spectral index as the VLA data and the same
  flux density at 1~keV as the Chandra data, to allow a comparison of
  X-ray and radio spectral indices.}
\end{figure*}
\begin{figure}
\plotone{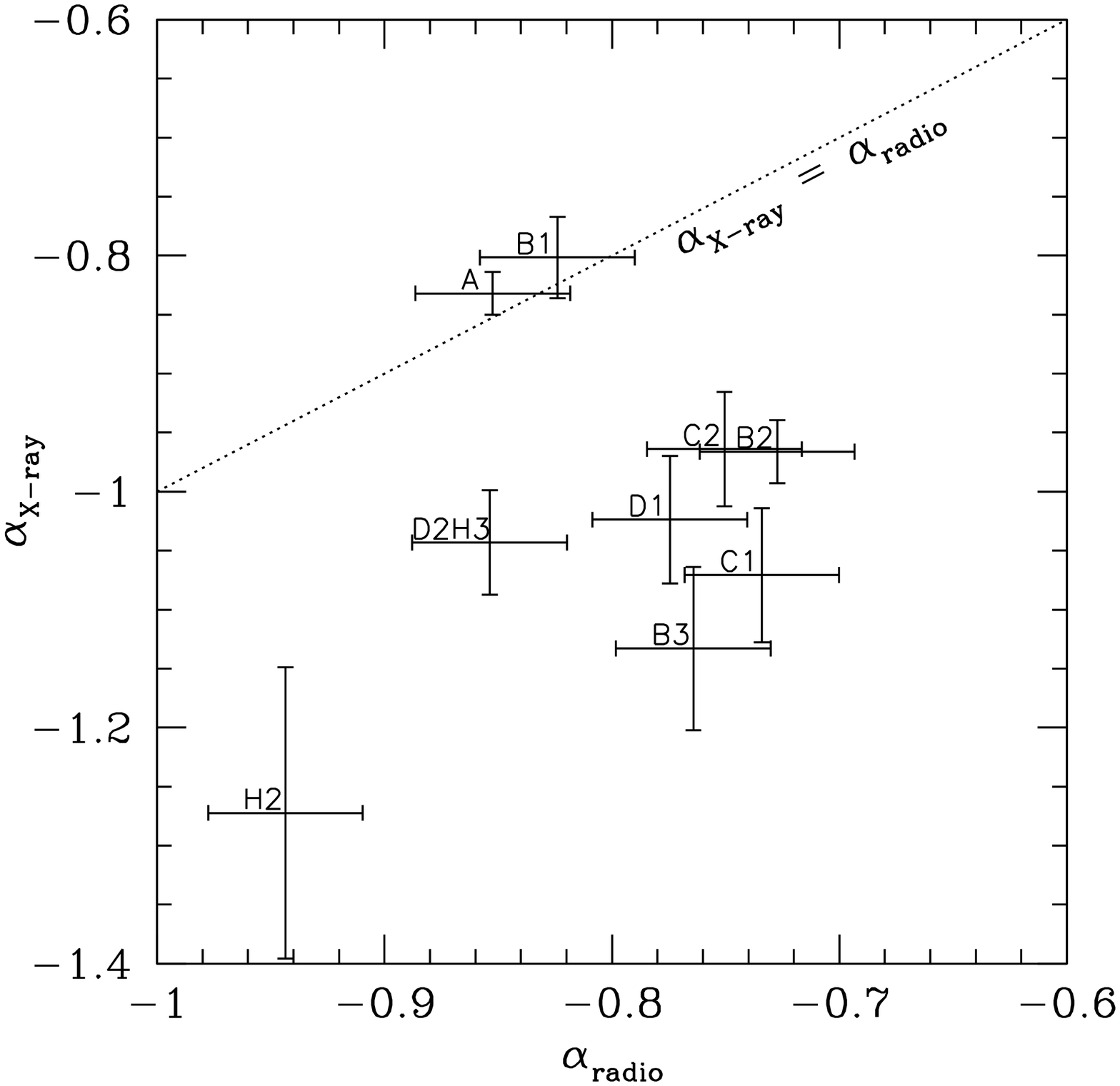}
\caption{\label{f:alpha_alpha}Comparison of radio and X-ray spectral
  indices for the jet regions from Fig.~\protect\ref{f:xraymap}.  The
  X-ray spectral index and its error are obtained from the Sherpa
  fit. The radio spectral index is the same least-squares fit shown in
  Fig.~\protect\ref{f:SEDs}, its error has been calculated from the
  signal-to-noise ratio of the VLA maps.  The dotted line shows
  equality of both spectral indices.  Except for regions A and B1, the
  X-ray spectrum is significantly softer than the radio spectrum.}
\end{figure}
Table~\ref{t:fits} gives the fit results.  In all cases, a simple
power law was a good fit to the data, according to the probabilities
obtained for the $\chi^2$ fits.  We also generated X-ray SEDs in six
energy bands from corresponding brightness maps. The SEDs for knots~A
and B2 show some curvature in the range 4--5keV and a slightly softer
spectrum at lower energies than the Sherpa fits, which would be
consistent with a marginally detected high-energy cutoff. However, no
evidence for a cutoff was found in the fits using the event lists,
neither in the pure power-law fits nor in fits with a cut-off power
law (Sherpa model \texttt{xscutoffpl}). The X-ray spectral indices of
A and B1 are consistent with each other and both approximately $-0.8$.
The knots in the remainder of the jet have slightly softer X-ray
spectra; all knots from B2 up to and including D2H3 again have
spectral indices that are consistent with each other and a value of
$-1.010 \pm 0.018$.

The radio hot spot region H2 is significantly detected in X-rays in
the combined data set; in fact, \citet{Mareta01} already noted that
the jet was detected out to 22\arcsec\ from the core, with the radio
peak in H2 being located 21\farcs4 from the core.  A comparison of the
X-ray flux profile of regions D2H3 and H2 with that of the nearly
point-like region~A confirms the reality of the detection of H2.  Its
spectrum is marginally softer than the rest of the jet.

Figure~\ref{f:SEDs} shows the spectral energy distributions (SEDs) of
the nine jet regions from radio up to X-rays. It is evident from the
SEDs that the jet emission from radio up to X-rays cannot be explained
as a single synchrotron component for \emph{any} part of the jet.
Except for region~A, this was already known from the analysis of ROSAT
data \citep{Roe00} and of the first Chandra data sets
\citep{Mareta01,Sam01}.  \citet{Jes02,JRMP05} showed that a
single-component synchrotron spectrum was not in fact viable for
knot~A, chiefly because the new multi-frequency multi-configuration
VLA data set used by them, and here, yielded substantially different
radio SED shapes.

Our spectral index measurement for knot~A of $\alpha= -0.83\pm 0.02$
is consistent with the 90\% confidence interval $\alpha_x =
1.1^{+0.5}_{-0.3}$ given by \citet{Sam01} at the $1.5\,\sigma$ level.
However, \citet{Mareta01} obtained a much harder X-ray spectrum with
$\alpha = -0.60 \pm 0.05$.  The discrepancy is due to two systematic
effects.  First, pileup slightly hardened the data used by
\citeauthor{Mareta01}. Secondly, corrections for ACIS contamination,
which absorbs X-ray photons mostly below 1~keV, were released only
after publication of their paper \citep[see][]{MTGea04}.  We can
reproduce the \citeauthor{Mareta01} result by turning off the pileup
and contamination correction.

In Fig.~\ref{f:alpha_alpha}, we compare the radio and X-ray spectral
indices directly.  It clearly shows the basic and important finding of
our spectral analysis: except for regions~A and B1, the X-ray spectrum
is significantly softer than the radio spectrum.

\section{Discussion: Implications of the new SEDs for the X-ray
  emission mechanism}
\label{s:discussion}

The aim of the Chandra observations presented here was a test of both
the synchrotron and the beamed IC-CMB models for the jet's X-ray
emission.  The non-detection of significant variability does
\emph{not} rule out synchrotron as X-ray emission process, as the
Chandra resolution element is sufficiently large (the FWHM of
0\farcs78 corresponds to 2.1\,kpc) to allow possible variability to be
washed out.  We now discuss the implications of the observed SED shape
for both synchrotron and beamed IC-CMB as X-ray emission mechanism.
The models have to account for the following observational facts:
\begin{enumerate}
\item \label{i:2spec} There is a spectral hardening of the
  near-infrared/optical/ultraviolet emission in all parts of the jet
  \citetext{see the SEDs in \citealp{Jes02} and Fig.~\ref{f:SEDs}, but
    in particular those in \citealp{JRMP05}}.  The shape of the second
  optical/UV emission component cannot be constrained to a high degree
  of certainty based on the present data set, but it may well be the
  extrapolation of the X-ray power law to lower frequencies
  \citep{Jes02}.
\item \label{i:xopt} The X-ray emission cannot be explained by a
  simple extrapolation of the total optical emission \citep[the
    ``bow-tie'' problem; see][]{Roe00,Sam01,Mareta01}.
\item \label{i:radioalpha} The radio synchrotron emission from the jet
  in 3C\,273 has a spectral index $\alpha \approx -0.75$ at all
  observed frequencies, down to 330\,MHz \citetext{\citealp{jetII} and
    confirmed by our unpublished VLA data}.
\item \label{i:pol} Radio and optical emission show very similar
  degrees of linear polarization out to 18\arcsec\ from the core, and
  the polarization vectors  are parallel to each other in all parts of
  the jet except for the optically quiet radio ``backflow'' or
  ``cocoon'' to the south of the jet \citep{RM91,jetIV}.
\item \label{i:xrad} The X-ray emission has a softer spectrum than the
  radio emission in most parts of the jet (Figures \ref{f:SEDs} and
  \ref{f:alpha_alpha}).
\item \label{i:xmorph} The same morphological features are seen at all
  wavelengths, from radio to optical \citep{Baheta95} to X-rays
  \citep{Sam01,Mareta01}.
\end{enumerate}
Items \ref{i:2spec} and \ref{i:xopt} individually rule out a one-zone
synchrotron model, but would be compatible with a two-zone synchrotron
model as well as an IC-CMB model with one or more zones.  Item
\ref{i:radioalpha} implies that the jet's total radio emission is
dominated by a single synchrotron component down to the lowest
currently observed frequencies.  The high linear polarization (item
\ref{i:pol}) lead to the conclusion that the optical emission is
synchrotron light \citep{RM91}, and the similarity of the radio and
optical polarization indicated that the radio and optical synchrotron
emission are due to the same electron population \citep{jetIV}.
Remarkably, if the X-ray emission could be shown to be of the same
origin as \emph{most}, and not just \emph{some}, of the optical
emission \citep[compare][]{Jes02}, the fact that the optical emission
is synchrotron would imply that the X-ray emission is synchrotron,
too; this might be expected from the similarity of the optical and
X-ray morphology (item \ref{i:xmorph} above), and is very strongly
supported by the analysis of Spitzer mid-infrared observations by
\citet{Uea06}.  The clarification of this issue requires the analysis
of further mid-infrared (Spitzer) and ultraviolet (HST) data. For now,
we discuss the viability of the IC-CMB and the two-zone synchrotron
models given item \ref{i:xrad}.

\subsection{Viability of the IC-CMB model}
\label{s:disc.IC}

\subsubsection{Single-zone IC-CMB}
\label{s:disc.IC.onezone}

In the beamed IC-CMB model, CMB photons are upscattered into the
Chandra band by electrons with Lorentz factors of order 50-200
\citetext{in a jet with bulk Lorentz factor $\Gamma$ and Doppler
  factor $\delta$, photons with incoming energy $E_\mathrm{cmb}$ are
  upscattered to energy $E_\mathrm{x}$ by electrons with Lorentz
  factor $\gamma \approx
  (E_\mathrm{x}/E_\mathrm{cmb}/[\Gamma\,\delta\,(1+z)])^{1/2}$, and we
  used $\Gamma = \delta = 15$, z = 0.158, $0.5\,\mathrm{keV} <
  E_\mathrm{x} < 8\,\mathrm{keV}$, $E_\mathrm{cmb} =
  6.6\times10^{-4}$\,eV; see \citealp{HK02}}.  Using the formula for
the characteristic frequency $\nu_\mathrm{c}$ of synchrotron emission
for electrons with Lorentz factor $\gamma$ in a magnetic field with
flux density $B$, $\nu_\mathrm{c}=
4.2\,\mathrm{MHz}\;B/(10\,\mathrm{nT})\,(\gamma/100)^2\,\delta\,/(1+z)$
($10\,\mathrm{nT} \mathrel{\widehat{=}} 100\,\mu$G), we obtain that these same electrons
produce synchrotron emission at frequencies 20--250\,MHz\,$\times
B/(10\,\mathrm{nT})$ i.e., at lower frequencies than the VLA
observations presented here \citep[the minimum-energy field in this
  jet is of order 10--20\,nT for $\delta=1$, and an order of magnitude
  smaller for $\delta=15$;][]{JRMP05,HK02,SSO03}.  The electron energy
distribution should be either a power law, or have curvature such that
lower-frequency emission has a harder spectrum.  Hence, in the beamed
IC-CMB model one expects that the X-ray spectrum should have a
spectrum that has the same spectral index as, or is harder than, the
radio emission from the jet.

As shown by Figures \ref{f:SEDs} and \ref{f:alpha_alpha}, the X-ray
spectrum is significantly softer than the radio spectrum in all parts
of the jet except for regions~A and B1.  When the beamed IC-CMB model
is put forward to account for X-ray emission from high-power radio
jets, or used to infer physical parameters such as the Doppler factor
and the line-of-sight angle, the usual assumption is a single-zone
single-component model, i.e., one in which the electron population
producing synchrotron emission is identical to that upscattering CMB
photons into the X-ray band
\citep[see][e.g.]{Tav00,Sam01,HK02,SGMea04,KS05,MSLea05} and where
identical jet volumes contribute to the synchrotron and
inverse-Compton emission.  This conventional single-zone beamed IC-CMB
model is difficult to reconcile with our new Chandra data for those
parts of the jet in which the X-ray spectrum is softer than the radio
spectrum, i.e., the entire jet from Region B2 outwards, at projected
distances greater than 15\arcsec\ from the core.

\subsubsection{Two-zone IC-CMB}
\label{s:disc.IC.twozone}

Compared to positing that the X-rays from jets in FR~II sources are
due to synchrotron emission, the single-zone beamed IC-CMB model
seemed to be more conforming to Occam's razor, since it only needed
the assumption that the bulk Lorentz factors inferred for parsec-scale
jets from VLBI observations of Blazars persist on the kiloparsec
scale.  By comparison, the synchrotron hypothesis would require at
least a two-zone model.  In fact, single-zone models are inadequate
for many jet observations \citep[e.g.,][]{HHSea99,HMW04,Cel00,JRMP05},
leading to the invocation of a second spatial zone and/or a second
spectral component.  Progress in distinguishing between the various
models can only be made if they are falsifiable by observations;
X-ray telescopes with spatial resolution comparable to that achieved
with radio and optical telescopes would be highly desirable in this
context.

\paragraph{A two-zone model from velocity shear}
The presence of a transverse velocity gradient, i.e., velocity shear,
in high-power jets would naturally create zones in the jet with
different inverse-Compton emissivities.  Based on high-resolution
observations and a physical model of the jet flow, the presense of
velocity shear has been established in a number of low-power jets:
those in the radio galaxies 3C\,31 \citep{LB04}, B2~0326+39 and
B2~1553+24 \citep{CL04} and NGC 315 \citep{CLBea05}.  A
``spine-sheath'' shear structure has also been implied for low-power
jets based on beaming statistics of FR~I and BL~Lacs in the B2 survey
\citep{CCCEa02}, as well as from polarimetric observations of the jet
in M87 \citep{Pereta99}.  However, both theoretical considerations
\citep{Meier02} and the analogy between accretion in AGN and black
hole X-ray binaries \citep{MHD03,FKM04} may lead us to expect that the
velocity structure of high-power jets is different from that of
low-power jets \citep[see the discussion in \S7 of][]{FBG04}.  While
the polarization properties of some jets in FR~IIs show a spine-sheath
structure \citep{SBB98,ARW99,PGVea05} that may suggest a similar
velocity structure as in FR~Is, \citet{PGVea05} have pointed out that
the polarization structure may be caused by the magnetic field
structure rather than a velocity gradient.  Thus, even though such a
structure has not yet been firmly established, it would not be
implausible for FR~II jets to have a spine-sheath structure,
warranting its consideration here.

\begin{figure*}
\plotone{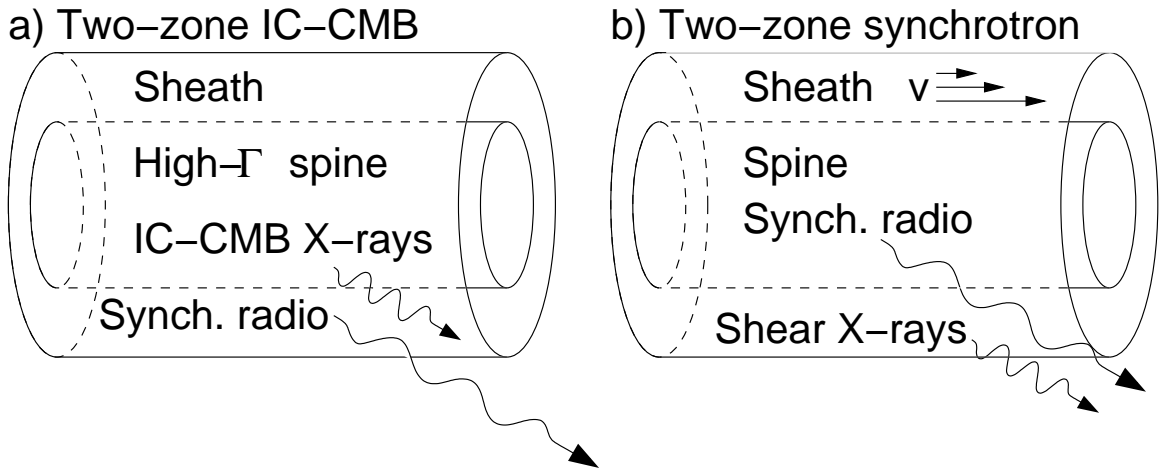}
\caption{\label{f:twozone}Illustration of two-zone emission models. In
  the two-zone inverse-Compton model (IC-CMB; a), the jet needs to
  have a highly relativistic spine ($\Gamma \approx 50-100$) to
  produce the X-ray emission via upscattering of cosmic microwave
  background photons without contributing significant radio
  synchrotron emission. The slower sheath produces the bulk of the
  radio synchrotron emission; it may or may not be moving
  relativistically itself.  In our two-zone synchrotron model (b),
  there is still a fast spine surrounded by a shear layer.  In this
  case, the bulk of the radio synchrotron emission is produced in the
  spine. In the sheath, the velocity shear allows particle
  acceleration to take place, producing X-ray emission also by the
  synchrotron process (labelled ``Shear X-rays''; shear acceleration
  may also occur in the IC-CMB model, but is not essential there).
  The two-zone synchrotron model would admit a much lower bulk Lorentz
  factor for the spine than the two-zone IC-CMB model.}
\end{figure*}
In a two-zone beamed IC-CMB model for 3C\,273, the highly relativistic
part of the flow (the ``spine'') would account for the bulk of the
X-ray emission, while the slower part of the flow (the ``sheath'')
would dominate the synchrotron emission (Fig.~\ref{f:twozone} a).
Thus, the observed radio spectral index of the jet is that of the
sheath, with $\alpha_{\mathrm{sheath}}\approx -0.75$.  To match the
observed X-ray spectral index, the fast spine would need to have a
synchrotron spectrum with $\alpha_{\mathrm{spine}} \approx -1$.  To
keep the observed radio spectral index unchanged, the spine would also
need to have a much lower radio luminosity.  As the inverse-Compton
model requires the spine to have a large Doppler factor, the radio
emission from the spine \emph{cannot} be ``Doppler-hidden'' by having
a Doppler factor much less than unity \citep[cf.\ the suggestion of
  such Doppler hiding making the jet in 3C\,353
  edge-brightened;][]{SBB98}.  Instead, the spine has to have a much
lower radio brightness than the sheath. If the spine has an
(observer-frame) flux density at 1.6\,GHz of 10\% of that of the
sheath, its steeper-spectrum contribution would not change the total
spectral index between 1.6\,GHz and 330\,MHz by more than 0.05, i.e.,
within the spectral index errors.  Hence, we consider a 10\%
contribution at 1.6\,GHz the maximum plausible flux density of the
spine that would keep it undetectable as a separate steep-spectrum
component.

\paragraph{Estimating the bulk Lorentz factor}
We can now estimate the bulk Lorentz factors within a 2-zone model
from the X-ray:radio ratio. The X-ray emission has to be produced in
the spine via upscattering of CMB photons by an electron distribution
that would contribute only about 1\% of the jet's total synchrotron
luminosity (where the latter is inferred from the radio flux).  We
just estimated that the spine could contribute at most 10\% of the
radio flux density at 1.6\,GHz; by extrapolating the X-ray flux
density to 1.6\,GHz using the observed spectral index, we obtain that
the X-ray:radio ratio for the spine \citep[$R(1)$ in the notation
  of][]{HK02} is of order 2000. This is much larger than the value
$R(1) \approx 1$ inferred from the jet's total radio emission
\citep{HK02}.  Following the method presented by \citet{HK02}, we can
infer the likely Doppler and Lorentz factors for the spine if we have
an estimate of the magnetic field.  If there is equipartition of
energy between the magnetic field and the particles in both the spine
and the sheath, we can use the scaling of the minimum-energy field
$B_\mathrm{min}$ with source luminosity $L$, volume $V$ and Doppler
factor\footnote{$\delta = [\Gamma (1-\beta\mu)]^{-1}$ is the
  relativistic Doppler factor for an object moving at speed $\beta c =
  (1-\Gamma^{-2})^{1/2} c$ at angle $\theta = \arccos \mu$ to the line
  of sight} $\delta$ to estimate the relative magnetic flux
densities. In this case, $B_\mathrm{min} \propto (L/V)^{2/7}
\delta^{-5/7}$ \citep[equation A7]{SSO03} is the appropriate scaling;
with our estimate that the spine has a total synchrotron luminosity of
about 1\% of that of the sheath, and assuming that the spine occupies
$1/2$ of the jet diameter and hence $1/3$ of the sheath's volume, the
magnetic field in the spine would still be $\delta\rel^{-5/7} \times$
20\% of the sheath's (where $\delta\rel = \delta\spi/\delta\sh$ is the
relative relativistic Doppler factor of spine and sheath
\citep{GK03b}). Hence, following \citet{HK02}, we obtain that the bulk
$\Gamma\spi$ would have to be of order 50--100 to produce $R(1) =
2000$.

\paragraph{Jet deceleration in a two-zone IC-CMB model}
The synchrotron emission from the sheath will appear boosted in the
frame of the spine, thus providing an additional seed photon field for
Compton scattering \citep{GTC05}.  Depending on the relative sizes and
speeds of spine and sheath, the energy density of the sheath photon
field as perceived in the frame of the spine could be comparable to
that of the cosmic microwave background.  If this was the case, the
boosted CMB would need to contribute only half of the seed-photon
energy density.  As the boosted CMB energy density scales like
$\Gamma^2$, the required bulk Lorentz factor would be reduced only by
about $1/\sqrt{2}$.  At the same time, the sheath electrons will
upscatter seed photons from the spine, creating an additional IC
component.  In the case of 3C\,273, the dominant contribution to the
synchrotron luminosity would arise in the sheath \citep[``layer'' in
  the terminology of][]{GTC05} rather than the spine as assumed for
the blazars considered by \citet{GTC05}.  This additional
``mutual-Compton'' (MC) scattering will produce detailed SEDs that are
quite different from those arising in a one-zone synchrotron + beamed
IC-CMB model.  As noted by \citet{GTC05}, the MC scattering could
cause a deceleration of the spine (an ``inverse Compton-rocket''
effect).  This might provide a more natural explanation for the
deceleration that is necessary to account for the changing X-ray:radio
ratios along some extragalactic jets, including 3C\,273's (see
\citealp{H06} and in particular \citealp{MJHea05} for the deceleration
of this jet implied by a one-zone beamed-IC model). The FR~I flow
modeling by \citet{LB04,CL04,CLBea05} shows that these jets
decollimate as they decelerate, presumably by entrainment, while no
sign of decollimation is observed in 3C\,273 and other high-power
jets.

However, as noted by \citet{GTC05}, the large number of free
parameters of this kind of two-zone model makes it very difficult to
obtain definite predictions for the expected shape of the SED.  The
most stringent observational constraints will be imposed by verifying
whether or not the radio spectrum shows any signs of the
steeper-spectrum component assumed to arise in the spine at lower
frequencies than observed so far, e.g., in the frequency range around
100\,MHz that will soon be accessible with the Low-Frequency Array
(LOFAR).  In any case, the two-zone IC-CMB model requires that the
``spine-sheath'' jet should develop very different low-energy electron
populations in the different parts of the flow.  This may seem
somewhat artificial if the velocity structure arises from simple
velocity shear, unless the shape of the electron energy distribution
in jets is governed entirely by a distributed acceleration mechanism,
rather than by shock acceleration in the innermost part of the jet
\citep[compare the discussion of ``shock-like'' versus ``jet-like''
  acceleration in][]{hs_II}.  The different energy distributions might
arise more naturally if the two-zone flow existed from the outset
\citep[e.g., as in the model by][]{SPA89}.  A detailed parameter study
is necessary to explore the range of plausible total SEDs (including
synchrotron emission, beamed IC-CMB, and MC scattering); this is
beyond the scope of this paper.

\paragraph{Constraints on two-zone models from the jet
  morphology} We will present a detailed study of the X-ray morphology
based on the new data in a separate paper; the most relevant finding
so far is that most of the jet is now clearly resolved transversely
even at the Chandra resolution.  However, the limit to the width of
the brightness peaks in knots~A and B2 is smaller than the measured
sizes in the optical and radio bands \citep{MJHea05}, where they are
clearly resolved \citep{Baheta95,Jes01,JRMP05}. This is incompatible
with the beamed IC-CMB model --- even in a decelerating one-zone flow
such as that considered by \citealp{GK04}, the optical emission should
be the most concentrated because the ``optical synchrotron'' electrons
have the shortest synchrotron loss scales.  Thus, even though the SED
of knot~A admits a one-zone IC-CMB model, its morphology does not.  In
general, the presence of unresolved knot emission superimposed on
resolved more diffuse jet emission compels us to consider two-zone
models (although not necessarily only spine-sheath models).

In summary, if FR~II jets have a spine-sheath structure, and the spine
is highly relativistic, there will naturally be two zones in the jet
with different IC-CMB emission.  The ``mutual-Compton'' scattering of
the radiation from the one zone by electrons in the other might
provide a natural explanation for the apparent deceleration of 3C\,273
implied by the changing X-ray:radio ratio \citep{GTC05,H06}.  However,
a two-zone IC-CMB model appears to require unrealistically large bulk
Lorentz factors for the spine of the jet, of order 50--100.  Moreover,
the two-zone IC-CMB model does not intuitively explain why the
emission in the X-ray band has a softer spectrum than that in the
radio band.

\subsection{Viability of the synchrotron model}
\label{s:disc.syn}

\begin{figure*}
\plotone{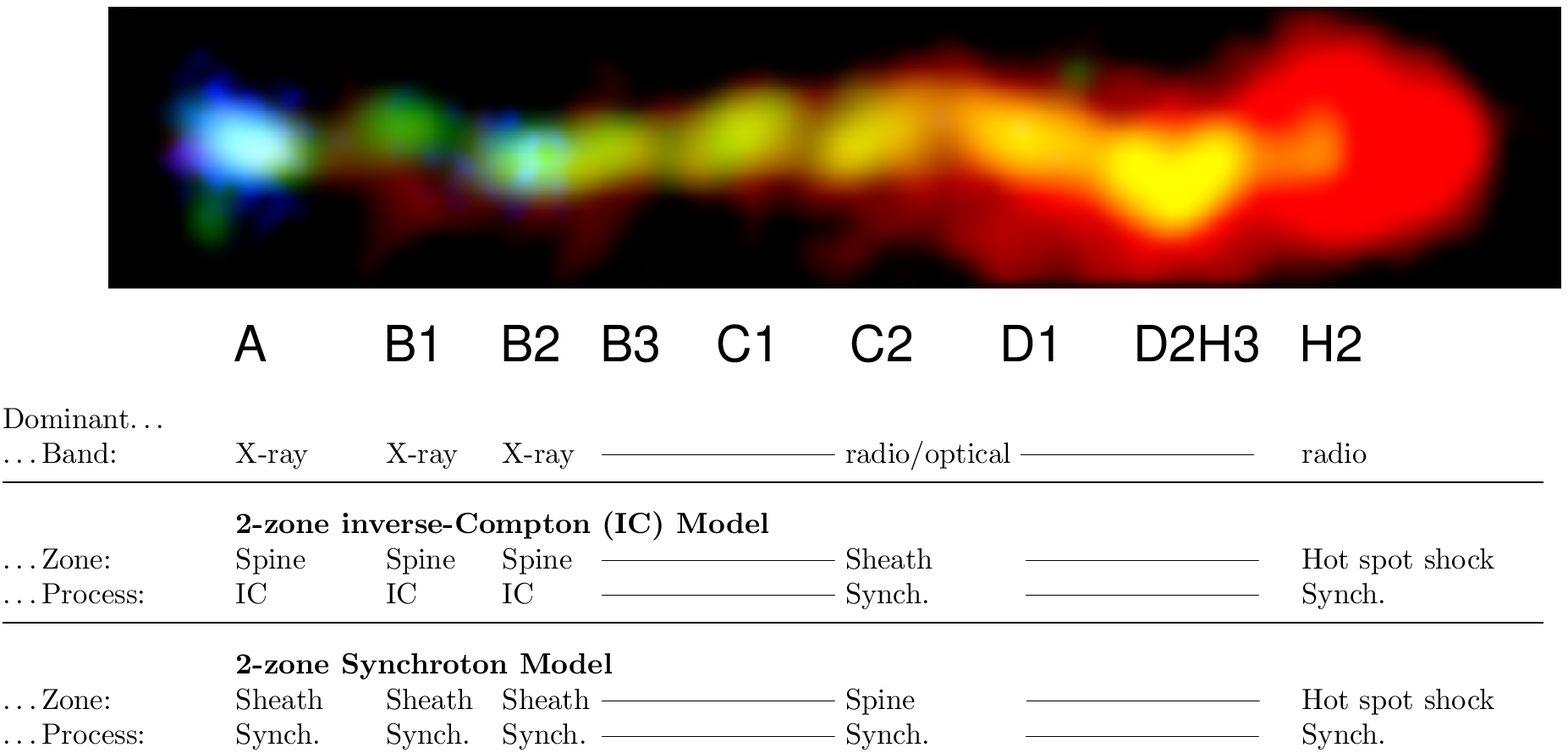}
\caption{\label{f:colorjet}False-color image of the jet in 3C273, made
  by sampling the individual images in Fig.~\protect\ref{f:xraymap} on
  a finer pixel grid, combining them to an RGB composite (blue
  channel: Chandra, green channel: HST, red channel: VLA) and
  smoothing slightly after combination to remove the sharp boundaries
  between the original pixels.  The energetically dominant waveband is
  given below the knots (compare Fig.~\protect\ref{f:SEDs}), as well
  as the dominant spatial component and corresponding emission
  mechanism in each of the two-zone, spine-sheath models considered in
  the text: a two-zone inverse-Compton model with a highly
  relativistic spine producing X-rays and some fraction of the optical
  emission via inverse-Compton scattering of CMB photons (IC-CMB),
  surrounded by a slower sheath producing radio synchrotron emission;
  and a two-zone synchrotron model in which the spine produces radio
  synchrotron emission, while the sheath is a shear layer that
  accelerates electrons emitting synchrotron X-rays.  In both models,
  knots A and B2 may have a contribution to the X-ray emission from a
  different emission or acceleration mechanism.}
\end{figure*}
As pointed out above, a one-zone synchrotron model is already excluded
by the spectral hardening in the near-infrared/optical/UV region.
Hence, we only discuss a two-zone synchrotron model here.  The
evidence in favor of a spine-sheath structure in FR~II jets summarized in
\S\,\ref{s:disc.IC.twozone} is of course equally relevant to the
discussion of a two-zone synchrotron model.

In a two-zone synchrotron model, there would be the same need to
account for these different electron energy distributions, but the
difference would now be between the \emph{low}-energy electron
distribution of one zone (emitting at radio wavelengths) and the
\emph{high}-energy electron distribution of the other one (producing
X-rays).  It is already known that many low-power jets emit
synchrotron X-rays \citep{WBH01}, and possibly \emph{all} such jets do
\citep{HWBea02}.  The extremely short synchrotron loss timescales of
10s of years for the X-ray emitting electrons require an
\emph{in-situ} acceleration process to generate these electrons.  It
would then be an obvious assumption that high-power jets are also able
to accelerate such electrons.  Indeed, some high-power jets show
evidence for synchrotron X-ray emission \citep{SGMea04,KHWM05,HC05}.
Moreover, \citet{SO02}, \citet{RieMan02} and \citet{RD04} have all
argued that efficient particle acceleration is possible in regions of
velocity shear.  \textbf{Thus, particle acceleration in a sheared
  region of the flow might provide a direct physical link between a
  two-zone velocity structure and a two-component synchrotron
  spectrum, since an additional acceleration mechanism would be
  available in the sheared part of the flow.}  We stress that the
presence of shear acceleration does not preclude the possibility that
there is another distributed acceleration mechanism acting in the
other, or both, fluid zones, e.g., one related to the dissipation of
magnetic fields \citep{Lit99}

In the context of such a model, the soft X-ray spectrum in 3C\,273
with $\alpha \approx -1$ for regions~B2 and beyond could be caused
either by a soft energy spectrum being generated by the distributed
acceleration process, or because it arises from a loss-dominated part
of the energy distribution.  The difference in X-ray spectral index
between the first bright knot~A and the faint bridge B1 on the one
hand, and the second bright knot B2 and beyond on the other hand,
would require at least some difference of the physical parameters
(either by producing a different electron energy distribution, or by
producing stronger cooling), and perhaps even a different acceleration
mechanism acting there.  This physical difference might have been
expected given that A is the first bright region of the optical and
X-ray jet, after a gap of 12\arcsec\ with only very faint optical
\citep{Marea03} and X-ray \citep{Mareta01} emission.  The difference
between A and B2 may well be related to the reason for the jet
lighting up at A.  We will revisit the emission from the inner part of
the jet in a forthcoming paper on the X-ray morphology.

In figure~\ref{f:colorjet}, we give a summary of the energetically
dominant wavebands and the associated spatial component and emission
mechanism, both for the two-zone inverse-Compton and the synchrotron
model. 

\section{Summary and outlook}
\label{s:conc}

We have presented new deep Chandra Acis-S3 observations of the jet in
3C\,273 and analyzed them in conjunction with archival data,
concentrating on the shape of the radio / optical / X-ray SED in this
paper.  Fitting power laws to the radio and X-ray spectra, we find
that a single-zone beamed IC-CMB model for the X-ray emission is
viable only in regions~A and B1; in the remainder of the jet, the
X-ray spectra are softer than the radio spectra.  This is in contrast
with the similarly well-studied X-ray jets of PKS\,0637$-$752,
1136$-$135, and 1150$+$497, where the radio and X-ray slopes are
compatible with each other in all parts of the jet where both have
been measured \citep{Cha00,SGDea06}.  Thus, most of the X-ray emission
from the jet in 3C\,273 must be due to a different emission process
than Compton scattering of CMB photons by the same electrons producing
the bulk of the synchrotron radio emission.

A single-zone synchrotron model extending from radio to X-rays had
already been ruled out for all parts of the jet based on previous
observations \citep{Roe00,Sam01,Mareta01,Jes02}.  A two-zone beamed
IC-CMB model does not seem capable of accounting for the X-ray
emission from 3C\,273's jet with any fewer assumptions than a two-zone
synchrotron model.  A two-zone IC-CMB model may still be viable, but
would require extreme bulk Lorentz factors to produce the observed
X-ray emission from much fewer electrons than inferred from the jet's
total radio synchrotron emission.  In addition, there is no intuitive
explanation for obtaining different electron energy distributions in
the spine and the sheath.  By contrast, a two-zone synchrotron model,
in which a distributed particle acceleration mechanism related to
velocity shear is producing the X-ray emitting particles, provides a
causal relationship between the difference in fluid velocities and the
different spectral properties.  There would still be velocity and
hence beaming differences between the two zones, but they would not
have to be as extreme as in the two-zone IC-CMB models, and therefore
one might hope to obtain better-constrained parameters for the two
zones, and hence falsifiable predictions of this model.  An X-ray
polarimeter would enable significant progress to be made in this
question; also, the mid-infrared wavelength region that has now been
made accessible by the Spitzer Space Telescope will be of great
importance, as shown by the analysis of this jet's mid-infrared
emission by \citet{Uea06}.

It would clearly be desirable to test the consistency of the radio and
X-ray spectral indices with the one-zone beamed IC-CMB model for more
jets in the future.  We will present a detailed analysis of the X-ray
morphology of 3C\,273's jet elsewhere \citep[for first results,
  see][]{MJHea05}.  This will allow more tests of both emission models
for the X-rays.  In particular, understanding how the (mostly)
wavelength-independent morphology of this and other jets is produced
will provide crucial insights into both emission and particle
acceleration mechanisms acting in this and other jets.

\acknowledgments

Work at SAO was partially supported by NASA contract NAS8-03060 and
grants NASGO3-4124A and NASGO4-5120C. HLM was supported from SAO
contract SV3-73016 to MIT for support of the Chandra X-Ray Center,
operated by SAO on behalf of NASA under contract NAS8-03060.  SJ
received support from NASA contract NASGO4-5120A, from the
U.S.\ Department of Energy under contract No.\ DE-AC02-76CH03000, and
from the Max-Planck-Gesellschaft through an Otto Hahn fellowship. He
thanks both the CfA and MIT for hospitality. We are grateful to Tom
Maccarone and especially Aneta Siemiginowska for providing essential
advice on spectral fitting, and to the anonymous referee for a
constructive report that helped in improving the presentation of our
results.




\end{document}